\documentclass[12pt]{article}
\usepackage[T1]{fontenc}
\usepackage{setspace}
\makeatletter

\newcommand{\LyX}{L\kern-.1667em\lower.25em\hbox{Y}\kern-.125emX\@}

\makeatother

\begin{document}

\begin{titlepage}

{\par\centering \textbf{\large The Euclidean travelling salesman problem: Frequency
distribution of neighbours for small-size systems} \vskip .2in \par}

{\par\centering \textbf{Anirban Chakraborti}\footnote{
\emph{email address} : anirban@cmp.saha.ernet.in
} \par}

{\par\centering \textit{Saha Institute of Nuclear Physics}, \textit{1/AF Bidhan
Nagar, Kolkata 700 064, India.}\par}

\noindent \vskip .3in 

\noindent \textbf{Abstract} 

\noindent We have studied numerically the frequency distribution \( \rho (n) \)
of the \( n \)-th neighbour along the optimal tour in the Euclidean travelling
salesman problem for \( N \) cities, in dimensions \( d=2 \) and \( d=3 \).
We find there is no significant dependence of \( \rho (n) \) on either the
number of cities \( N \) or the dimension \( d \). 

\noindent \vskip 2in 



\noindent \end{titlepage}

\noindent \newpage 

\noindent \textbf{1. Introduction} 

\noindent The study of optimization problems is of considerable interest to
computer scientists, mathematicians and physicists alike. In a typical optimization
problem, there is a large finite set of possibilities to search from, in order
to obtain the optimal solution: if the problem is of ``size'' \( N \), then
typically, there are of the order of \( N! \) or \( e^{N} \) possibilities,
of which we want the one that minimizes (or maximizes) the cost function. The
travelling salesman problem (TSP) is a simple example of a combinatorial optimization
problem where, given a certain set of cities and the intercity distance metric,
a travelling salesman must find the shortest tour in which it visits all the
cities and returns to its starting point \cite{BHH,G}. It is a non-deterministic
polynomial complete (NP-complete) problem. There are two forms of TSP which
are of interest: the Euclidean TSP and the random link TSP. In the Euclidean
TSP, the \( N \) cities are distributed with uniform randomness in a \( d \)-dimensional
hypercube and the distance is measured in the Euclidean metric (with \( \Delta l=\sqrt{\Delta x^{2}+\Delta y^{2}} \)),
whereas in the random link TSP, the distances between the cities \( i \) and
\( j \) are taken as independent random variables with a given distribution.
It was noted that the random link TSP can be mapped onto the Euclidean model,
provided the distribution is chosen appropriately and the correlations between
three or more distances neglected \cite{MPV}. 

A city is said to be the \( n \)-th neighbour of a reference city if there
are exactly (\( n-1 \)) other cities that are nearer to the reference city.
In a given configuration of cities, for every city we can find its neighbours,
arrange them in order of their distances from that city and label them consecutively
with their neighbour number \( n \). Thus \( n=1 \) is the nearest neighbour,
etc. We can find out how many times the \( n \)-th neighbour is chosen along
the optimal tour and determine the frequency distribution \( \rho (n) \). Here,
we have studied numerically the frequency distribution \( \rho (n) \) along
the optimal tour for the Euclidean TSP only. The optimum tours are obtained
with the help of branch and bound algorithms with open boundary conditions.
As a matter of interest, we have also studied the frequency distribution \( \rho (n) \)
using the ``greedy algorithm''.

\vskip .2in

\noindent \textbf{2. Simulation and numerical determination of} \emph{\( \rho (n) \)}

We generated random configurations in dimensions \( d=2 \) and \( 3 \), for
different sizes \( N=10 \) to \( 256 \). For each configuration, we determine
the frequency distribution of neighbours \( \rho (n) \) along the optimal tour
and then take the average over \( 5000 \) configurations. We have found the
optimum tours in \( d=2 \) and \( 3 \), with the help of branch and bound
algorithms using open boundary conditions (Fig. 1). The results for \( \rho (n) \)
in dimensions \( d=2 \) and \( 3 \) are shown in Fig. 2 (a) and (b). The numerical
values of \( \rho (n) \) for different values of \( n \) in \( d=2 \), are
plotted against \( 1/N \) in Fig. 2 (c). These show that the frequency distribution
\( \rho (n) \) does not vary significantly with \( N \) and so \( \rho (n) \)
does not have any prominent finite-\( N \) effect, and it does not depend on
the dimension \( d \). The frequency distribution of neighbours, obtained by
numerical fitting, is of the form \( \rho (n)=A\exp (-an)[1+B\exp (n/b)] \)
where \( A=0.72\pm 0.01 \), \( a=0.60\pm 0.03 \), \( B=0.05\pm 0.01 \) and
\( b=3.0\pm 0.1 \). The errors in \( A \), \( a \), \( B \) and \( b \)
are obtained by eye-estimation. 

As a matter of interest, we have also studied \( \rho (n) \) using the ``greedy
algorithm'', where from each city the salesman goes to the nearest city not
already in the tour and finally from the \( N \)-th city returns directly to
the first. The results for \( \rho (n) \) in dimensions \( d=2 \) and \( 3 \)
are shown in Fig. 3. Here also, \( \rho (n) \) does not depend on the dimension
\( d \). There seems to be a cross-over from the exponential decay to a power
law decay at \( n=6 \), for some unapparent reasons.

\vskip .2in

\noindent \textbf{3. Discussions and summary}

It was shown in \cite{BHH} that the average optimal travel distance in the
unit \( d \)-dimensional hypercube, where \( N \) random cities are distributed
uniformly, \( <l_{N}^{(d)}> \) scales as \( N^{1-1/d} \). The empirically
obtained frequency distribution \( \rho (n) \) along the optimal tour can be
used to get a rough estimate of this average optimal travel distance \( <l_{N}^{(d)}> \).
We may write \( <l_{N}^{(d)}>\simeq \sum ^{N-1}_{n=1}<\rho (n)><D_{N}^{(d)}(n)> \)
where \( <...> \) denote the ensemble averages and \( <D^{(d)}_{N}(n)> \)
is the average \( n \)-th neighbour distance along the optimal tour between
\( N \) cities in a unit \( d \)-dimensional hypercube. Note that the average
optimal travel distance \( <l_{N}^{(d)}> \) should actually involve the average
over the product \( <\rho (n)D_{N}^{(d)}(n)> \), while we use here the product
of averages \( <\rho (n)><D_{N}^{(d)}(n)> \). Here, we use the general expression
given in \cite{M} for the \( n \)-th neighbour distance in the unit \( d \)-dimensional
hypercube containing \( N \) random points distributed uniformly: \( <D_{N}^{(d)}(n)>=\left[ \Gamma (d/2+1)/(\pi ^{d/2}N)\right] ^{1/d}\left[ \Gamma (n+1/d)/\Gamma (n)\right]  \),
and the empirically determined frequency distribution \( \rho (n) \) (\( =<\rho (n)> \))
to estimate \( <l_{N}^{(d)}> \). We find that for \( d=2 \), \( <l_{N}^{(2)}>\simeq 0.77N^{1/2} \)
and for \( d=3 \), \( <l_{N}^{(3)}>\simeq 0.74N^{2/3} \). It may be noted
that the ensemble of \( D_{N}^{(d)}(n) \) in TSP is not the same as the case
where distances are calculated without any restrictions.

In summary, here we study numerically the frequency distribution \( \rho (n) \)
along the optimal tour for the Euclidean TSP for dimensions \( d=2 \) and \( 3 \).
For optimum tours, we find there is no significant dependence of \( \rho (n) \)
on either the number of cities \( N \) or the dimension \( d \), and the empirically
determined frequency distribution \( \rho (n)\sim A\exp (-an)[1+B\exp (n/b)] \).
Since \( \rho (n) \) does not have any prominent finite-\( N \) effect and
the values of \( \rho (n) \) remain significant for small \( n \) (up to \( n\sim 45 \)
in our study for \( N\leq 256 \)), one can therefore determine \( \rho (n) \)
quite accurately using small system sizes, and thus optimizing the computational
efforts. 

\vskip 0.2in

\noindent \textbf{Acknowledgement} : The author is indebted to Prof. B. K. Chakrabarti
for introducing the problem and for useful discussions, and is grateful to G.
Chatterjee and N. Dutta for their support and encouragement.

\noindent \newpage

\noindent \textbf{Figure captions}

\noindent \vskip .1 in

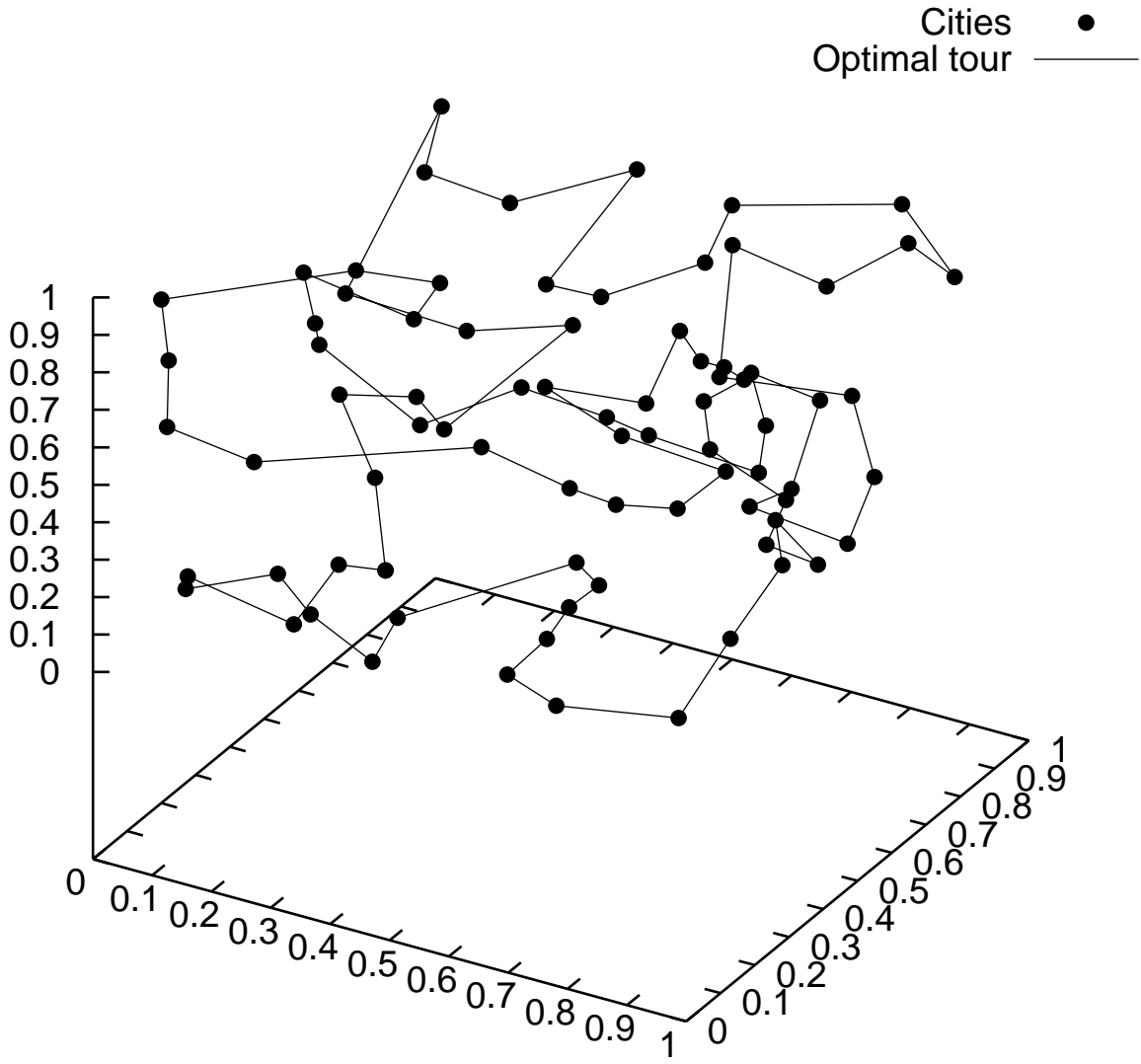
\noindent \textbf{Fig. 1} : \textbf{(a)} A typical optimal tour is shown for
\( N=80 \) cities in a unit area for dimension \( d=2 \). The cities are represented
by black dots and the optimal tour is indicated using a solid line \textbf{(b)}
Similarly, a typical optimal tour is shown for \( N=80 \) cities in a unit
volume for dimension \( d=3 \). The cities are represented by black dots and
the optimal tour is indicated using a solid line.

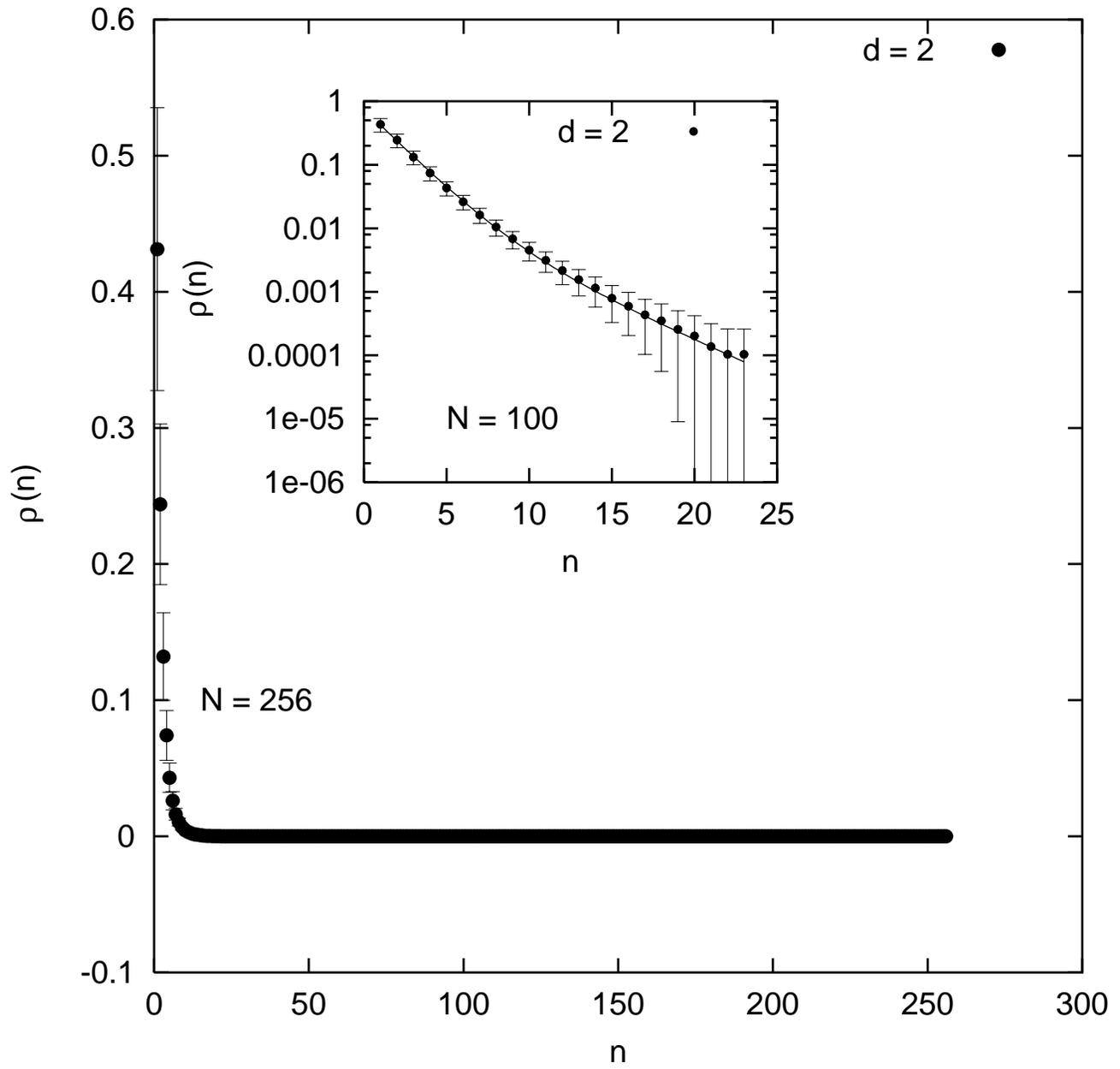
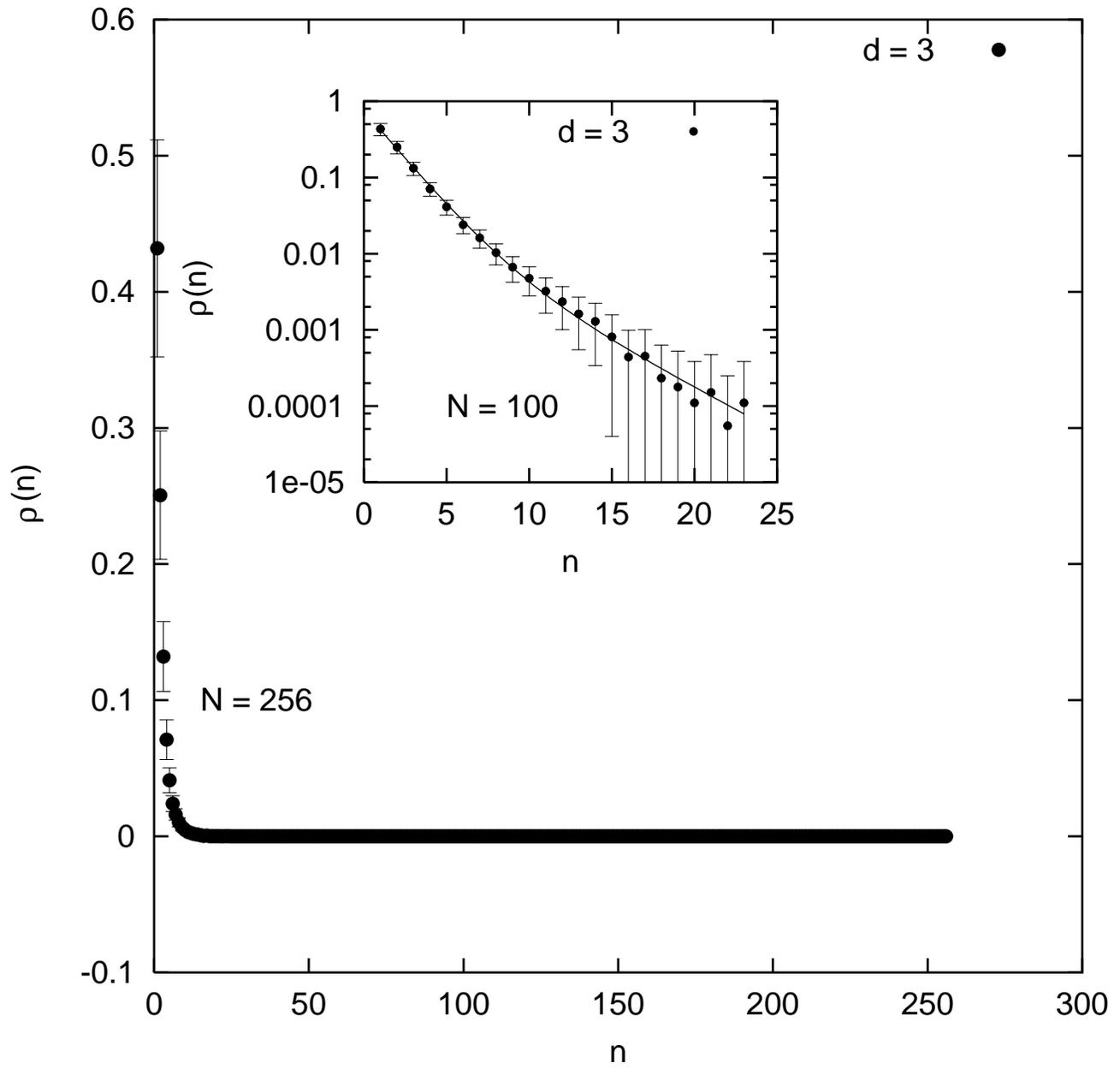
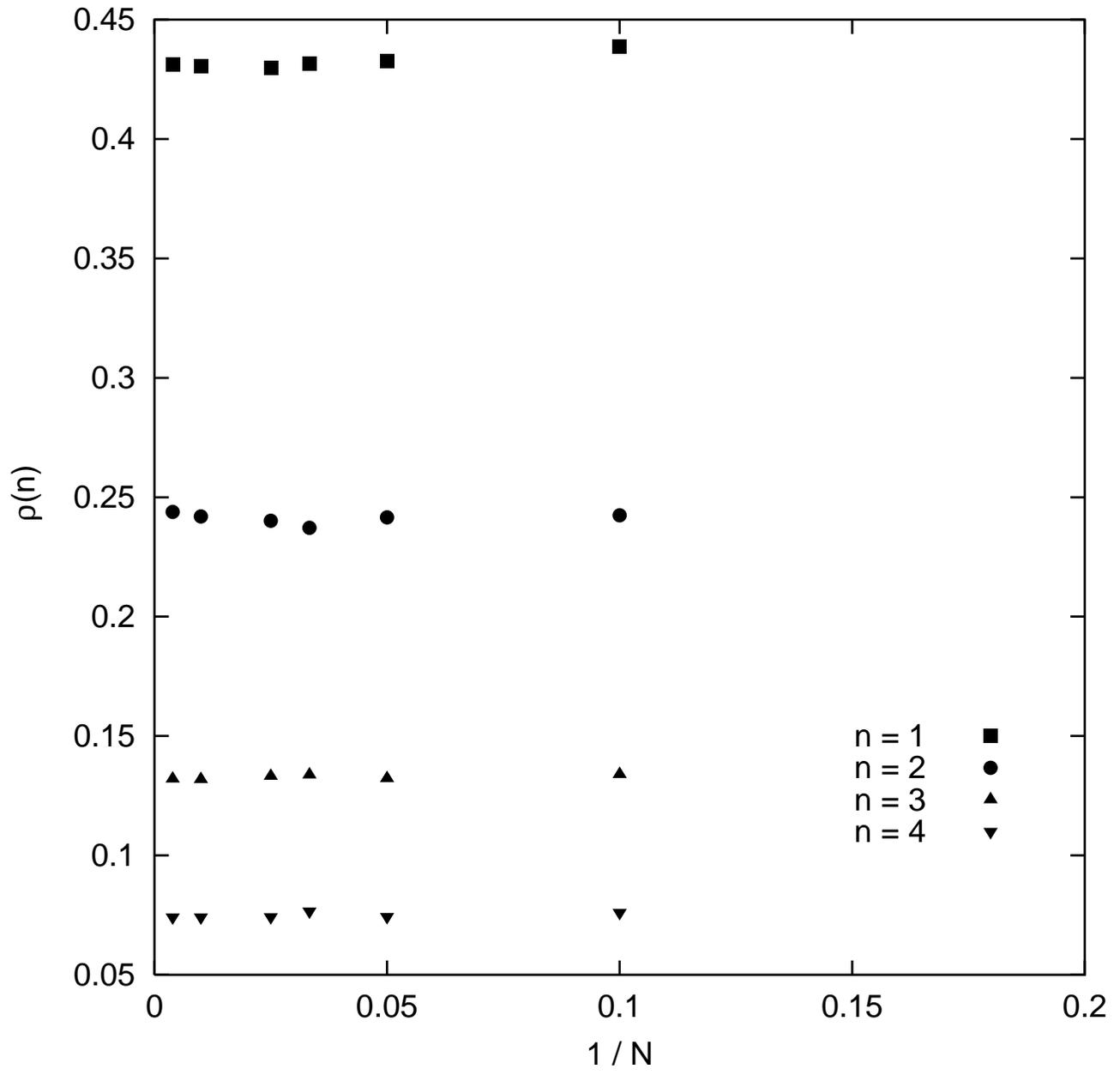
\noindent \textbf{Fig. 2} : \textbf{(a)} Plot of the distribution \( \rho (n) \)
of neighbours on an optimal tour of \( N=256 \) cities, for \( d=2 \). In
the inset, \( \rho (n) \) for \( N=100 \), in \( d=2 \) is plotted in the
linear-log scale and the numerically fitted curve \( \rho (n)=0.72\exp (-0.60n)[1+0.05\exp (n/3.0)] \)
is shown by a solid line; the error bars are due to configurational fluctuations
\textbf{(b)} Plot of the distribution \( \rho (n) \) of neighbours on an optimal
tour of \( N=256 \) cities, for \( d=3 \). In the inset, \( \rho (n) \) for
\( N=100 \), in \( d=3 \) is plotted in the linear-log scale and the numerically
fitted curve \( \rho (n)=0.72\exp (-0.60n)[1+0.05\exp (n/3.0)] \) is shown
by a solid line; the error bars are due to configurational fluctuations \textbf{(c)}
Values of \( \rho (n) \) for different values of \( n \) are plotted against
\( 1/N \) for \( d=2 \), to show that the frequency distribution does not
have any significant finite-\( N \) effect in the range \( 10\leq N\leq 256 \)
considered.

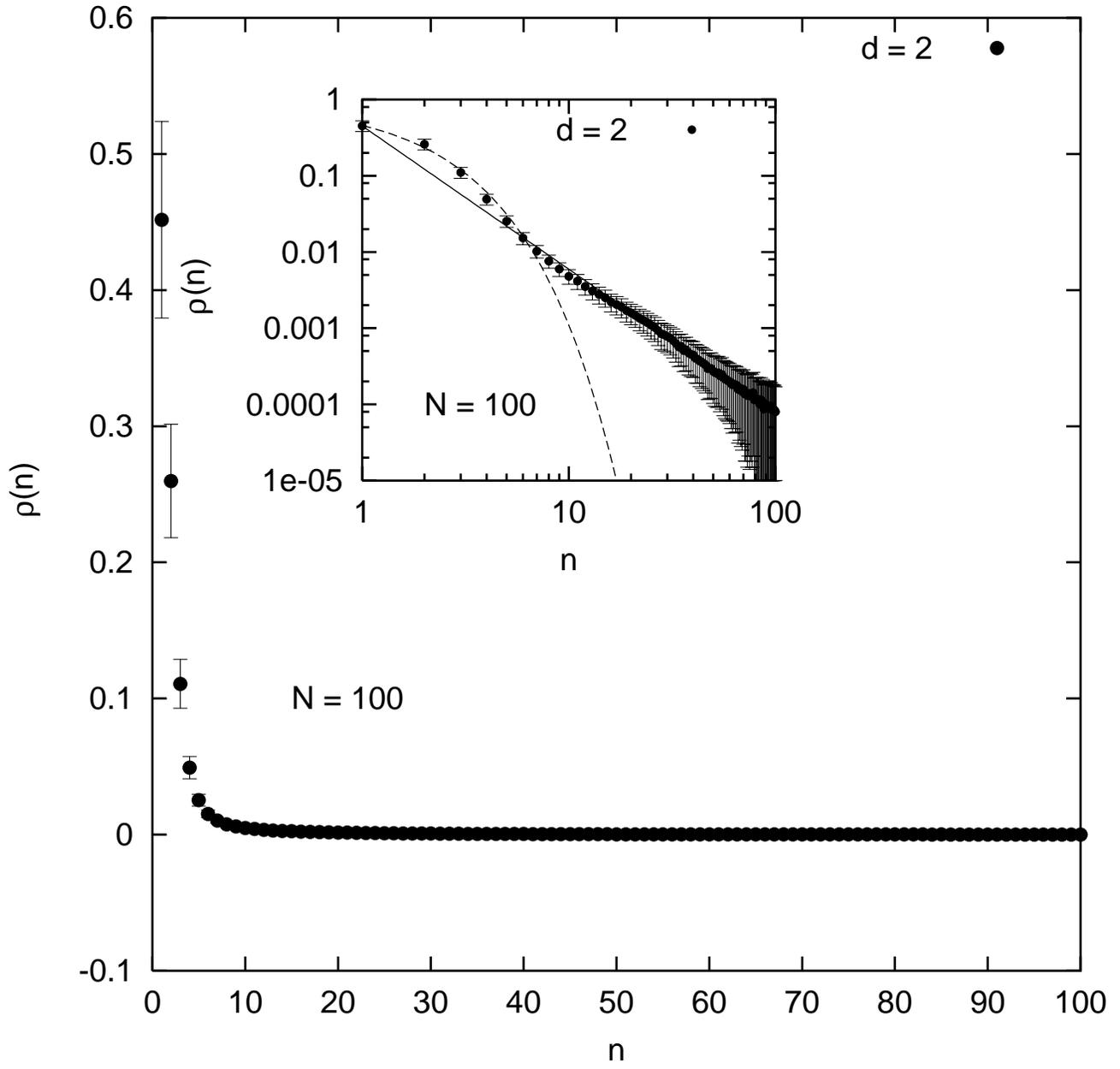
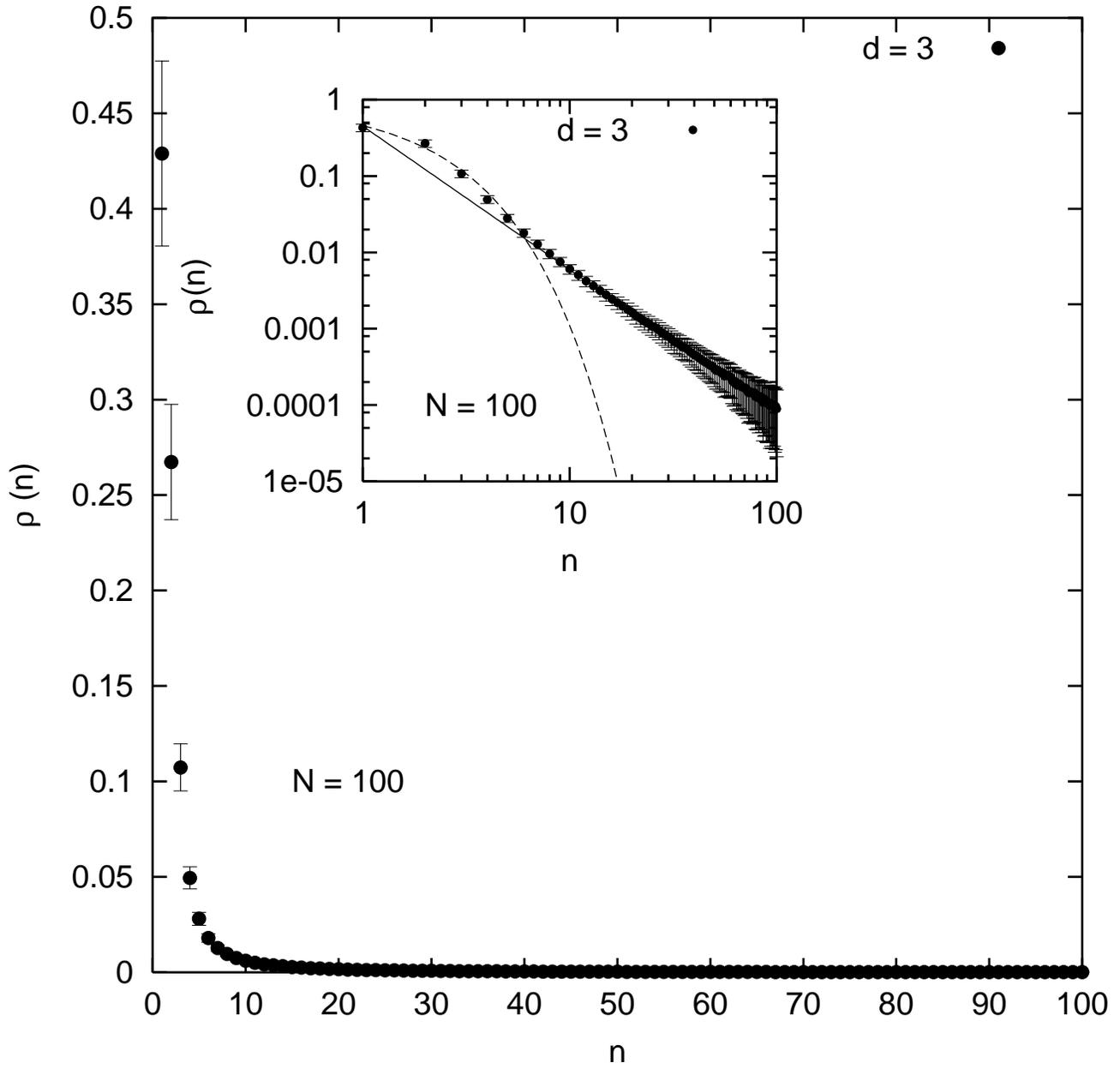
\noindent \textbf{Fig. 3} : \textbf{(a)} Plot of the distribution \( \rho (n) \)
of neighbours on a ``greedy'' tour of \( N=100 \) cities, for \( d=2 \). In
the inset, \( \rho (n) \) for \( d=2 \) is plotted in the log-log scale and
the numerically fitted curves are shown by a dashed line (exponential decay)
and a solid line (power law decay); the error bars are due to configurational
fluctuations \textbf{(b)} Plot of the distribution \( \rho (n) \) of neighbours
on a ``greedy'' tour of \( N=100 \) cities, for \( d=3 \). In the inset, \( \rho (n) \)
for \( d=3 \) is plotted in the log-log scale and the numerically fitted curves
are shown by a dashed line (exponential decay) and a solid line (power law decay);
the error bars are due to configurational fluctuations. 

\end{document}